\documentstyle[preprint,aps,floats]{revtex}

\begin{document}

\title{Isotopic Dependence of the Casimir Force}

\author{Dennis E. Krause$^{1,2}$ and Ephraim Fischbach$^{2}$}

\address{$^{1}$Physics Department, Wabash College, Crawfordsville, IN 
47933-0352}
\address{$^{2}$Physics Department, Purdue University, West Lafayette, IN 
47907-1396}

\date{\today}

\maketitle

\begin{abstract}
We calculate the dependence of the Casimir force on the isotopic composition of the
interacting objects.  This dependence arises from the subtle influence of the nuclear
masses on the electronic properties of the bodies.  We discuss the
relevance of these results to current experiments utilizing the iso-electronic effect to
search  at very short separations for new weak forces suggested by various unification
theories.

\end{abstract}

\pagebreak

The Casimir effect \cite{Casimir} has been the subject of intense study recently, both
experimentally and theoretically \cite{Bordag Review,Books,Lamoreaux,Mohideen
experiments,Harris,Ederth,Chan,Bressi}.  Not only is it of interest in its own right, as a
novel and fundamental quantum effect, but the increasing precision of experimental tests
of the Casimir effect has led to its use in searching for new macroscopic forces acting
over sub-millimeter scales
\cite{Bordag 1998,Most and Novello 2001,Fischbach 2001a}.  This
follows from the fact that the Casimir force becomes the dominant background to any new
macroscopic force over short distance scales after  electrostatic and magnetic effects have
been eliminated.  If we consider, for example, the Casimir force per unit area
$P_{C}(d)$ between two perfectly conducting parallel plates at temperature $T = 0$
separated by a distance
$d$, then \cite{Casimir}
\begin{equation}
			P_{C}(d) = -\frac{\pi^{2}\hbar c}{240}\frac{1}{d^{4}}
         = -\frac{0.013}{(d/\mbox{$\mu$m})^{2}}\,\mbox{dyn/cm$^{2}$}.
\label{Casimir pressure}
\end{equation}
It is instructive to compare $P_{C}(d)$ to the gravitational force:  For two Cu plates
having dimensions 1 cm$^{2} \times 1$ mm a distance $d$ apart, the Casimir force exceeds
the Newtonian gravitational force when $d \lesssim 14$ $\mu$m.  (This distance becomes
slightly larger at room temperature
\cite{Krause}.)  However, this is the very distance scale over which recent
extra-dimensional theories suggest the possibility of new short-range gravitational
forces \cite{Arkani,Randall,Randall Review}.  If follows that detecting such forces
requires understanding the (presumably larger) Casimir background.

For Casimir experiments using real (rather than idealized) conductors, the effects of
finite conductivity, finite temperature, and surface roughness become important, and
considerable progress has been made in recent years in dealing with these effects
\cite{Bordag Review,KlimMost 2001,Bezerra,Lambrecht}.  Still, the inherent difficulty in 
calculating the Casimir force for real materials to high precision limits the use of
Casimir force experiments to constrain new  forces.
The object of the present paper is to implement a recent proposal
\cite{Fischbach 2001a,Krause,Fischbach 2001b} which aims to sidestep these problems
when searching for new very short-ranged macroscopic forces.  The approach 
relies on the fact that the Casimir force depends primarily on the {\em electronic}
properties of the interacting bodies, while the proposed new forces, including
those arising from new spatial dimensions,  depend on the test bodies'
{\em nuclear}, as well as on their  electronic, properties.  One should therefore be able
to set limits on new forces at sub-micron separations by measuring the {\em differences} in
forces between test bodies composed of different isotopes of the same element (the
iso-electronic effect), since the Casimir force should be independent of isotope to a good
approximation.  Any observed differences between the test bodies could then be attributed
to new physics after other effects (differences in sample preparation, etc.) have
been accounted for.  Furthermore, this method does not require a detailed
calculation of the Casimir force,  if this force  is known to
be the same for the isotopes being compared.   However, before one can extract reliable
limits on new forces from an experiment  based on the iso-electronic effect, one must be
confident that any small isotopic dependence of the Casimir force will produce a force
difference that is less than the force resolution of the experiment.  
In what follows we present the first calculation of the isotopic dependence of the Casimir
force.  This calculation is of interest for two reasons:  First, it is directly relevant
to an experiment currently underway \cite{Decca} to search for new short-range forces by
comparing the Casimir forces for two isotopes of the same element.  Secondly, our results
reveal new characteristics of the Casimir force, the dependence on lattice spacing and
nuclear masses, which may be utilized in future applications.

To understand how the isotopic dependence of the Casimir force comes about we note that
for two infinitely thick parallel plates composed of real dielectrics at $T \neq 0$, the
expression for the Casimir force $F_{C}(d,T)$ is more complicated than Eq.~(\ref{Casimir
pressure}), and is given by the Lifshitz formula \cite{Lifshitz 1956,Lifshitz 1961,Lifshitz
1980,KlimMost 2001}:
\begin{eqnarray}
F_{C}(d,T) & = & -\frac{k_{B}TA}{\pi c^{3}}\sum_{l=0}^{\infty}\!^\prime\,
\xi_{l}^{3}\int^{\infty}_{1}p^{2}dp
\nonumber\\ 
&  &
\times \left\{\left[\left(\frac{K(i\xi_{l},T) + \varepsilon(i\xi_{l},T)p}
       {K(i\xi_{l},T) - \varepsilon(i\xi_{l},T)p}\right)^{2}
      e^{2d(\xi_{l}/c)p} - 1\right]^{-1}\right.
       \nonumber\\ 
 & & \nonumber\\ 
&  &  \phantom{sp}\mbox{} + \left.\left[\left(\frac{K(i\xi_{l},T) + p}
       {K(i\xi_{l},T) - p}\right)^{2}
      e^{2d(\xi_{l}/c)p} - 1\right]^{-1}\right\}.
\label{Lifshitz formula}
\end{eqnarray}
Here $A$ is the area of the plates,  $k_{B}$ is Boltzmann's constant, $\omega = i\xi_{l}$,
$\xi_{l} = 2\pi k_{B}Tl/\hbar$, and
\begin{equation}
K(i\xi_{l},T) = \left[p^{2} - 1 + \varepsilon(i\xi_{l},T)\right]^{1/2},
\end{equation}
where $\varepsilon(\omega,T)$ is the frequency and temperature dependent dielectric
constant. The prime on the summation sign indicates that the $l~=~0$ term should be
multiplied by 1/2.
The dielectric properties of the interacting media determine the Casimir
force acting between real metallic plates.  
In practice, one obtains $\varepsilon(\omega,T)$ from a Drude or plasma model for
low frequencies and from tables of optical data for higher frequencies.

We note from Eq.~(\ref{Lifshitz formula}) that the Casimir force depends on the
temperature $T$, as well as on $d$, and that this $T$-dependence between real metals
enters in two different ways
\cite{Bezerra}.  First, the quantum electromagnetic field is at temperature $T$ and this
contributes a thermal pressure from real photons on the plates which becomes
significant for larger plate separations.  When $d \gtrsim \hbar c/2k_{B}T$,
the energy spacing between the field modes decreases allowing thermal energy to more
easily excite higher modes.  [Note that for a room temperature experiment ($T = 300$ K),
$\hbar c/2k_{B}T = 3.8\,
\mbox{$\mu$m}$, while for an experiment operating at liquid helium temperatures ($T = 4$
K), $\hbar c/2k_{B}T = 0.29\,$ mm.] If
$d
\ll
\hbar c/2k_{B}T$, there is a sufficiently large gap between the ground state and first
excited states to prevent significant thermal excitation of higher modes in
which case
$F(d,T)$ reduces to \cite{Lifshitz
1956,Lifshitz 1961,Lifshitz 1980,KlimMost 2001}
\begin{eqnarray}
F_{T}(d) & = & -\frac{\hbar A}{2\pi^{2} c^{3}}
\int^{\infty}_{0}d\xi\, \xi^{3}\int^{\infty}_{1}p^{2}dp
\nonumber\\ 
&  &
\times \left\{\left[\left(\frac{K(i\xi,T) + \varepsilon(i\xi,T)p}
       {K(i\xi,T) - \varepsilon(i\xi,T)p}\right)^{2}
      e^{2d(\xi/c)p} - 1\right]^{-1}\right.
       \nonumber\\ 
 & & \nonumber\\ 
&  &  \mbox{} + \left.\left[\left(\frac{K(i\xi,T) + p}
       {K(i\xi,T) - p}\right)^{2}
      e^{2d(\xi/c)p} - 1\right]^{-1}\right\}.
\label{Lifshitz  T = 0 formula}
\end{eqnarray}
As was observed recently \cite{Bezerra},  we can see from
Eq.~(\ref{Lifshitz T = 0 formula}), that there remains another temperature
dependence to the Casimir force.  The dielectric constant
$\varepsilon(\omega,T)$ is also temperature dependent as discussed below, even
when one can neglect the temperature fluctuations of the field.  Therefore,
following Ref.~\cite{Bezerra}, we define $F(d,T= 0) \equiv F_{T}(d)$, were the
subscript denotes this implicit $T$-dependence.  Thus,
it is important to check that the tabulated data for the dielectric constant are
appropriate for the temperature at which the experiment is conducted.  This
has not been a problem in previous experiments since they  have all been conducted
at room temperature at which most of the tabulated optical data are obtained. 
However, these data may be inappropriate for experiments conducted at low
temperatures.

For the case of two infinite plates, we see that the isotopic dependence of the
Casimir force must enter through $\varepsilon(i\xi,T)$, which  depends mainly upon the
electronic properties of the material.   Optical data for the isotopes of interest, at 
temperatures relevant for an experiment, are difficult to find in the literature. 
Furthermore, experience from room temperature Casimir force experiments indicates that,
ideally,  one should obtain optical data directly from the actual samples used, since there
is sufficient variation from sample to sample.  

In the absence of relevant experimental data, we can estimate the isotopic
dependence of the Casimir force from theoretical considerations.  As we have discussed,
the Casimir force is determined through the Lifshitz formula by the dielectric constant
$\varepsilon(\omega,T)$.   For metals which can be described by the plasma model,
$\varepsilon(\omega,T)$ is  characterized by a single parameter, the plasma frequency
$\omega_{p}$ which depends, in turn, on the lattice constant $a$.  In the presence of an
anharmonic potential, the lattice spacing $a$ will be different for two isotopes of the
same element, since the zero point motion of the isotopes at $T = 0$ depends on the
respective isotopic masses.  Thus, the isotopic dependence of the Casimir force arises from
the dependence of the lattice constant on mass, and this dependence affects the dielectric
constant, and eventually the Casimir force, through the Lifshitz formula.

To quantify the preceding discussion, we consider a Casimir force
experiment utilizing conductors which can be described by the plasma model.  Although this
is a simple model of metals, it is sufficiently reliable for our present purposes.  In the
plasma model the dielectric constant $\varepsilon(\omega = i\xi)$ is given by
\begin{equation}
\varepsilon(i\xi) = 1 + \frac{\omega_{p}^{2}}{\xi^{2}},
\label{plasma model}
\end{equation}
where the plasma frequency $\omega_{p}$ is
\begin{equation}
\omega_{p}^{2} = \frac{4\pi Ne^{2}}{m_{\rm eff}V}.
\label{plasma frequency}
\end{equation}
Here $N/V$ is the number of conduction electrons/volume and $m_{\rm eff}$ is the
effective electron mass.    If  Eq.~(\ref{plasma model}) is substituted into
Eq.~(\ref{Lifshitz  T = 0 formula}), one finds in the limits $d \gg 2\pi c/\omega_{p}$ and
$T \ll
\hbar c/2k_{B}d$ \cite{Hargreaves,Bordag Review},
\begin{equation}
F(d) \simeq -\frac{\pi^{2}}{240}\frac{\hbar c A}{d^{4}}
  \left(1 - \frac{16}{3}\frac{c}{\omega_{p}d}\right).
\label{Plasma Casimir}
\end{equation}
In the simplest case, let 
$N_{\rm val}/V$ be the number of {\em valence} electrons/volume and let $m_{\rm eff} =
m_{e}$, the free electron mass, in which case, Eq.~(\ref{plasma frequency}) reduces to 
\begin{equation}
\omega_{p}^{2} = \frac{4\pi N_{\rm val}e^{2}}{m_{e}V}.
\label{simple plasma frequency}
\end{equation}
From Eq.~(\ref{simple plasma frequency}), we see that in this case, all of the isotopic
dependence must arise from $V$, the volume per atom, which is proportional to $a^{3}$. 

The isotopic dependence of the lattice spacing has been a topic of interest for some time
\cite{London,Plekhanov}.  It is well-known that the temperature dependence of $a$, which
leads to thermal expansion of solids, arises from anharmonic terms in the interatomic
potential
\cite{Kittel}.  For example, in a one-dimensional lattice with a typical interatomic
distance given by $x$, let an  atom's potential energy be approximated by 
$V(u) \simeq (1/2)ku^{2} - (1/6)bu^{3}$,
where  $u = x - x_{0}$, $x_{0}$ is the equilibrium separation, $k$ is the effective spring
constant, and
$b$ characterizes the anharmonic contribution.  At temperature
$T$, the thermal average displacement depends on the anharmonic term so that the lattice
constant in this model becomes temperature dependent and proportional to $b$
\cite{Kittel}:
\begin{equation}
 a(T)= x_{0} + \langle u\rangle \simeq x_{0} + \frac{b}{2k^{2}}k_{B}T,
\end{equation}
where $k_{B}$ is Boltzmann's constant.  This temperature dependence of the lattice
spacing affects the plasma frequency, which leads to a temperature-dependent dielectric
constant $\varepsilon(\omega,T)$ as mentioned earlier.  If one replaces thermal vibrations,
which allow atoms to sense the anharmoniticity of the potential, with quantum zero-point
motion ($kT
\rightarrow \hbar\omega/2$), one finds that the lattice spacing in this model becomes 
dependent on the atomic mass $M$, since the vibrational frequency $\omega$  is
proportional to
$1/\sqrt{M}$ \cite{Ramdas}:
\begin{equation}
 a(T = 0) \simeq x_{0} +  \frac{b}{4k^{2}}\hbar\omega .
\end{equation}
It follows that the temperature and isotopic dependence of
$\varepsilon(\omega,T)$ are linked through the anharmonic term of the interatomic
potential.  At finite temperatures, both thermal and zero-point motions contribute,
although the former dominate at higher temperatures. Hence the isotopic dependence
of $a$ is most significant at temperatures much less than the Debye temperature. 
For a current review of theoretical estimates and experimental results for the isotopic
dependence of the lattice constant, see Plekhanov \cite{Plekhanov}. 
With the exception of nickel, which is one of the metals being considered for an
experiment utilizing the iso-electronic effect \cite{Fischbach 2001a}, most of this effort
has focused on materials which would not be appropriate for Casimir force experiments. 
Nonetheless,  one finds (Table
\ref{isotope table}) that  $\Delta a/a \sim 10^{-4}$ for those elements that have been
studied, and this agrees with theoretical estimates.

If we assume that the entire isotopic dependence of the Casimir
force is dominated by the isotopic dependence of the lattice constant through
Eq.~(\ref{simple plasma frequency}), and that the Casimir force is given by
Eq.~(\ref{Plasma Casimir}), then we find that a variation of the lattice constant $\Delta
a$ leads to a  relative difference in the Casimir force for two different
isotopes,
\begin{equation}
\frac{\Delta F_{21}}{F} \simeq 
  \left(\frac{16}{3}\frac{c}{\omega_{p}d}\right)\frac{\Delta\omega_{p}}{\omega_{p}}
= 
-\left(\frac{8c}{\omega_{p}d}\right)
\frac{\Delta a_{21}}{a},
\end{equation}
where $\Delta F_{21} = F_{2} - F_{1}$, $\Delta a_{21} = a_{2} - a_{1}$, and we have
used $\Delta\omega_{p}/\omega_{p} = -(1/2) \Delta V/V = -(3/2)\Delta a/a$.   Since
Eq.~(\ref{Plasma Casimir}) is valid  when
$2\pi c/\omega_{p}d
\ll 1$, and experimentally  one finds (e.g., for nickel) $\Delta a/a \sim 10^{-4}$, one
expects 
\begin{equation}
\frac{\Delta F_{21}}{F} \ll 10^{-4},
\end{equation}
under these conditions. This is several orders of
magnitude below the current resolution of Casimir force experiments ($\Delta
F/F \sim 10^{-2}$).  However, since this problem remains unexplored experimentally, it may
be possible to find situations in which the isotopic $\Delta F/F$ is large enough to be
observed. 

To summarize, we have shown that for metals which can be described by the plasma model,
the relative difference in the Casimir force between plates composed of different isotopes
separated by $d \gg 2\pi c/\omega_{p}$ is negligible in any current experiment
\cite{Decca} utilizing force differences to extract limits on new forces.  For metals
which are not well described by the plasma model, and for experiments with shorter
separations, further analysis will be needed to ascertain how well our conclusions hold,
particularly when more realistic models of the dielectric constants are used for the actual
experimental samples.  Additionally, other effects of an isotopic mass difference should
be explored.  These include the isotopic dependence of $m_{\rm eff}$, and the possibility
that the dielectric constant can depend on isotopic mass in other ways besides the lattice
constant.  However, these effects are not likely to alter the principal conclusion of our
analysis, that current searches for new short-range forces using the iso-electronic effect
\cite{Decca} can ignore the isotopic dependence of the Casimir force.

\acknowledgements
The authors thank  G. Carugno, R. Decca, A. Lambrecht, D. L\'{o}pez, V. M. Mostepanenko,
A. W. Overhauser, A. K. Ramdas, S. Reynaud, G. Ruoso, and S. Rodriguez for helpful
discussions. This work was supported in part by the U. S. Department of Energy under
contract No. DE-AC02-76ER071428.


\pagebreak

\begin{table}
\caption{Experimental values of $\Delta a/a$ for several isotopes.}
\label{isotope table}
\begin{tabular}{lcc}
Isotopes & $\Delta a/a$  & Reference \\ \hline\hline
$^{58}$Ni, $^{64}$Ni & $1.4 \times 10^{-4}$ [$T$ = 78 K] & 
\cite{Kogan 1962} \\
 & $5.7 \times 10^{-5}$ [$T$ = 300 K] & 
\cite{Kogan 1962} \\ \hline
$^{12}$C, $^{13}$C (Diamond) & $-1.5 \times 10^{-4}$ [$T$ = 298 K] &
\cite{Holloway} \\ \hline
$^{6}$Li, $^{7}$Li & $-2 \times 10^{-4}$ [$T$ = 293 K] & \cite{Covington}
\\ \hline
$^{20}$Ne, $^{22}$Ne & $-1.9 \times 10^{-3}$ [$T$ = 3 K] & 
\cite{Batchelder} \\
&$-1.6 \times 10^{-3}$ [$T$ = 24 K] & 
\cite{Batchelder} \\ \hline
$^{70}$Ge, $^{76}$Ge & $-5.3 \times 10^{-5}$ [$T$ = 30 K] & 
\cite{Sozontov} \\
& $-2.2 \times 10^{-5}$ [$T$ = 300 K] & 
\cite{Sozontov} \\ 
\end{tabular}
\end{table}

\end{document}